\documentclass[runningheads]{llncs}
\usepackage[T1]{fontenc}

\usepackage{graphicx}
\usepackage{tikz,xcolor,hyperref,xurl}
\usepackage{comment}
\usepackage{subcaption}
\usepackage{tabularx}
\usepackage{tcolorbox}
\usepackage{amsmath}
\usepackage{amssymb}

\definecolor{lime}{HTML}{A6CE39}
\DeclareRobustCommand{\orcidicon}{%
	\begin{tikzpicture}
	\draw[lime, fill=lime] (0,0) 
	circle [radius=0.16] 
	node[white] {{\fontfamily{qag}\selectfont \tiny ID}};
	\draw[white, fill=white] (-0.0625,0.095) 
	circle [radius=0.007];
	\end{tikzpicture}
	\hspace{-2mm}
}

\foreach \x in {A, ..., Z}{%
	\expandafter\xdef\csname orcid\x\endcsname{\noexpand\href{https://orcid.org/\csname orcidauthor\x\endcsname}{\noexpand\orcidicon}}
}

\usepackage{graphicx}
\usepackage{tikz,xcolor,hyperref,xurl}
\usepackage{comment}
\usepackage{subcaption}
\usepackage{tabularx}
\definecolor{lime}{HTML}{A6CE39}
\DeclareRobustCommand{\orcidicon}{%
	\begin{tikzpicture}
	\draw[lime, fill=lime] (0,0) 
	circle [radius=0.16] 
	node[white] {{\fontfamily{qag}\selectfont \tiny ID}};
	\draw[white, fill=white] (-0.0625,0.095) 
	circle [radius=0.007];
	\end{tikzpicture}
	\hspace{-2mm}
}

\foreach \x in {A, ..., Z}{%
	\expandafter\xdef\csname orcid\x\endcsname{\noexpand\href{https://orcid.org/\csname orcidauthor\x\endcsname}{\noexpand\orcidicon}}
}

\begin{document}
\title{Star Network Motifs on X during COVID-19 \thanks{The research for this paper was supported by the following grants: Cognizant Center of Excellence Content Moderation Research Program, Office of Naval Research (Bothunter, N000141812108) and Scalable Technologies for Social Cybersecurity/ARMY (W911NF20D0002). The views and conclusions are those of the authors and should not be interpreted as representing the official policies, either expressed or implied.}}
\titlerunning{Star Motifs}
%
\author{Lynnette Hui Xian Ng*\inst{1}\orcidB, Divyaansh Sinha*\inst{1}\orcidA,
Kathleen M. Carley\inst{1} \orcidC}

\authorrunning{Ng et. al.}
\institute{Carnegie Mellon University, Pittsburgh PA 15213, USA 
\email{\{huixiann,divyaans,carley\}@cs.cmu.edu} \\ *Joint first authors}
\maketitle 
\begin{abstract}
Social network motifs are recurring patterns of small subgraphs that indicate fundamental patterns of social communication. In this work, we study the simple star network motifs that recur on X during the COVID-19 discourse. We study the profile of the manifestation of the star network among bot and human users. There are six primary patterns of the star motif, differentiating by the bots and humans being either egos and alters. We describe the presentation of each of these six patterns in our data, demonstrating how the motif patterns can inform social media behavioral analysis.

\keywords{network science \and graph motifs \and star \and social media bots}
\end{abstract}
\section{Introduction}
Social media platforms are extensively used for communication. The communication network between agents on social media can be represented as a graph $G=(V,E)$, where the vertice set $V$ are the agents, and the edge set $E$ are communication links. Within the social network graph consists of motifs, recurring patterns of small subgraphs that represent the building blocks of complex networks \cite{milo2002network}. Network motifs are a powerful analytical tool for understanding complex social phenomena. A common social network motif is the triad, a three-node pattern used to analyze relationships reciprocity, transitivity, and balance \cite{Wasserman_Faust_1994}. While triadic structures have received extensive attention to explain social relationships \cite{ruiz2023triadic}, other structures like the star motifs have been less systematically studied, especially in the context of social media interactions.

Star network motifs are motifs where a central ego node is connected to multiple peripheral nodes. These structures have long been known to be key elements in networks, and are important in a number of ways for understanding social capital distribution, and indicate pathways of information diffusion and influence. Foundational understanding of network structures traces back to pioneering work on sociometry by Moreno \cite{moreno1934whoshallsurvive}, which revealed star-like structures in group dynamics. Bavela's study of communication patterns in groups identified centralized star networks as efficient for problem-solving tasks \cite{bavelas1950communication}. Freeman provided the mathematical foundation for identifying central actors and associated metrics (e.g., the degree centrality metric), that are critical for identifying star structures \cite{freeman1979centrality}.
Recent work on egocentric networks show that these star-like structures are ubiquitous in social media, with egos typically maintaining a small core of close relationships (peripherals) surrounded by larger number of weaker connections ($nth$-degree connections) \cite{iniguez2023universal}. Work on network individualism notes how individuals maintain personal networks across digital and physical spaces, emphasizing the shift from group-based to person-centered connectivity \cite{wellman2002internet}.

But do all stars operate the same way? As noted by Simmel, central actors in networks, like the center of a star, may act in different ways \cite{simmel1950sociology}. Research using ego-networks has shown that similarity among alters, and similarity of alters with ego, can change the way the network operates and the effective power of the ego \cite{wellman2002internet}. This prior research, however, used networks where all the nodes are human, and so had the same affordances in terms of ability to communicate \cite{wellman2002internet}. We ask: \textbf{What about star networks that have bots as well as humans?}

The social media landscape has two key types of agents: bots and humans. Bots are cyber social agents that are pre-programmed to automatically interact on the social media platform \cite{ng2025global}. The distinction between human and bot behavior in network structures have become increasingly important for understanding information diffusion patterns. Profiling communication mechanics between bots and humans through star motifs revealed the social structures and patterns of social organization between organic and inorganic agents \cite{alrhmoun2023emergent}.

The process of information diffusion in a social network is inherently dependent on the structure of social relations. The influence effect is greatest for strong ties, peripheral nodes to a central node, or node pairs that are constantly communicating with each other \cite{bakshy2012role}. Such network effects have been shown to be a critical piece in persuading people online \cite{leung2014persuasion}. Further, these effects can differ depending on bot and human interaction mechanics \cite{khaund2021social}. Despite the recognition of the importance of network motifs in the understanding in social media dynamics, there remains a gap in how different agent types (i.e., bots vs humans) manifest within motif structures. We extend past systematic analysis of triadic motifs and properties to the star motif structure \cite{Wasserman_Faust_1994}, and characterize the interplay between bots and humans.

This work studies star network motifs in the communication networks derived from X during the 2022 COVID-19 pandemic, and profiles the presence of bots versus humans within the star network motifs. In 2020, the social media pandemic discourse had been characterized as an ``infodemic", with heightened attention to social media manipulation, misinformation amplification and growing awareness of coordinated inauthentic behavior \cite{cinelli2020covid}. From a taxonomy of six primary patterns of star motifs, we show, through a case study, the role of network motifs in the amplification of social communication. 

\section{Data Description and Processing}
For this study, we use data from \cite{blane2023analyzing}. This data was collected from X. These discussions took place a week before the newly developed coronavirus vaccine was officially launched, and was collected with vaccination-related keywords such as \#covidvaccine.  This dataset only accessed publicly available tweets and user accounts, and no attempt was made to access private accounts and tweets. We urge the readers to refer to \cite{blane2023analyzing} for detailed information about the data collection parameters. Within the collected data, there are 580,135 unique X users. 

\paragraph{Bot Identification}
Among the X users, we identify bot agents using the BotHunter algorithm. The BotHunter algorithm has a tiered random forest architecture. This algorithm is chosen because it was $\sim$90\% during its methodology building, and can efficiently work on offline datasets, such as the one we have on hand \cite{beskow2018bot}. BotHunter returns a probability score $P(bot)\in[0,1]$ for each agent. A score closer to 1 indicates that the agent is more bot-like.  We differentiate agents based on a
large-scale study on thresholding $P(bot)$ \cite{ng2022stabilizing}:
\begin{equation}
    Agent\_Type =
    \begin{cases}
      Bot & P(bot) \geq 0.7\\
      Human & P(bot) < 0.7 \\
      Human & \text{otherwise}
    \end{cases}  
\label{eq:isbot}
\end{equation}

After thresholding the $P(bot)$ scores, the BotHunter algorithm identified 153,379 (26.43\%) bot users. Other studies on social media bots estimate a 2-8\% of conservative/liberal bots \cite{chang2022comparative}, or an 18.5\% of general bots \cite{ng2025global}. It is plausible that our estimate is higher, because our dataset had been collected around the COVID vaccine discourse, a topic highly penetrated by bot narratives \cite{broniatowski2018weaponized}. 

\subsection{Retweet network construction}
Retweet networks capture information sharing and content propagation. The act of retweeting represents both implicit endorsement and information amplification, making these networks valuable indicators of influence dynamics \cite{boyd2010tweet}.

From our dataset, we construct a retweet network graph $G$, in which $G=(V,E)$. In a retweet network graph, the edge $e_{i,j}$ between $v_i$ and $v_j$ meant that $v_j$ retweeted $v_i$. Each edge has a weight $w_{e_{i,j}}$ that indicates the number of retweets between the two agents. A thicker edge indicates sustained retweeting actions by $v_j$ of $v_i$.

We use the ORA software\footnote{https://netanomics.com/ora-pro/} to visualize the structure of the retweet network (\autoref{fig:summary}). We prune the initial network to remove links with a weight less than 3 in order to focus on sizable agents with more meaningful relationships.

\subsection{Star Motifs from Ego Networks}
Ego networks describe the connections of an agent (the ego $n_0$) with its social peers (the alters $n \in N,\; n \neq n_0$) \cite{arnaboldi2017online}. Ego networks of the ego $n_0$ can be represented as one-hop graphs $G=(N,E)$. The one-hop network means that the alters $n_a = n_1...n_n$ are directly connected to the ego $n_0$ by an edge. Particularly, the edgeset $E = \{(n_i, n_0): 1 \leq i \leq n\}$ is where all the edges are oriented from $n_i$ to $n_0$ representing that alter $n_i$ retweets posts from the ego $n_0$. The structural properties of ego networks reflect many aspects of human social behavior \cite{arnaboldi2017online}, such as cooperation and information flow. 

In this paper, we focus on star-shaped network motifs patterns that appear in the ego networks. Star network motifs are defined in Definition 1, which, simply put, is the motif where one ego node connected to multiple peripheral alters. The alters should not have connections with each other, but since we were dealing with social media data, we relax this constraint and allow for the alters to have minimal links with each other. For simplicity, we studied the motifs with undirected links, which means that edges $E$ do not have a defined direction \cite{mao2020ramsey}.

\begin{tcolorbox}[
    colback=gray!5!white,
    colframe=gray!75!black,
    title= Definition 1: Star Network Motif for Social Networks,
    fonttitle=\bfseries,
    label=def:star_motif
]
A star motif $S_k$ is a connected subgraph with $k+1$ nodes consisting of one ego node $v_c$ and $k$ peripheral alters $\{v_1, v_2, \ldots, v_k\}$, where:
\begin{align*}
S_k &= (V, E) \text{ where } V = \{v_c, v_1, v_2, \ldots, v_k\} \\
E &= \{(v_c, v_i) : i \in \{1, 2, \ldots, k\}\}
\end{align*}
\textbf{Constraints:}
\begin{itemize}
    \item $\deg(v_c) = k$ (ego node has degree $k$)
    \item $\deg_{A}(v_i) \leq 2, \forall i \in \{1, 2, \ldots, k\}$ (each alter can have at most 2 edges with other alters)
    \item $\deg(v_i) \leq 3, \forall i \in \{1, 2, \ldots, k\}$ (each alter has degree at most 3)
\end{itemize}
\end{tcolorbox}

Star motifs are a basic form of a centralized network structure, and captures scenarios where the ego serves as a hub for multiple disconnected alters. Star motifs create natural information bottlenecks, because the central node controls the information flow between alters, giving the ego significant power through the structural advantage \cite{albert2002statistical}.



\section{Network Metrics of Bots vs Humans}
Network metrics create signatures for the two bot/human agent types, and serve as complementary features towards motif formation. From the retweet graphs, we calculate node-level network metrics that indicate the influence an agent has within the network. These metrics are: (1) Betweenness centrality, which is the extent that a node lies on the shortest path of other pairs of nodes in the network. The betweenness centrality indicates the extent that the node serves as an intermediaries node that controls the flow of information within the network. (2) Eigenvector centrality, that measures the nodes influence based on its connections to its neighbors. A high eigenvector centrality indicates that the node is more connected to other highly influential nodes, and that the nodes have access and influence over other important nodes in the network; and (3) Total Degree centrality, which measures the total number of direct connections of the nodes. This measure indicates the extensiveness of the direct reach of the node. 
We compare these network metrics between bots and humans using the student t-test with Bonferroni correction.

\autoref{tab:network_metrics} compares the statistics of the network metrics between bots and humans. There is a significant difference between the metrics for the total degree centrality, indicating distinct behavioral patterns between bots and humans in network engagement. This aligns with previous research where bots either aggressively form connections, or remain relatively isolated within their own botnets \cite{orabi2020detection}, both having differing total degree centrality than humans. This means that motifs with bot egos will differ with human egos in the number of connections.

The lack of significant differences in betweenness and eigenvector centrality suggested that while bots may differ from humans in terms of raw connectivity, bots do not necessarily occupy more strategically important positions within the network. Therefore, both bots and humans are equally likely to be egos of network motifs.


\vspace{-0.5cm}

\begin{table}
    \centering
    \begin{tabular}{|p{3cm}|p{2cm}|p{2cm}|p{2cm}|p{2.8cm}|}
    \hline
        \textbf{Metric} & \textbf{t-statistic} & \textbf{p-value} & \textbf{Corrected p-value} & \textbf{Significant ($p<0.05$) After Correction?} \\ \hline
    Betweenness Centrality & 2.65 & 8.09E-3 & 9.70E-2 & No \\ \hline 
    Eigenvector Centrality & 1.41 & 1.59E-1 & 1.00 & No \\ \hline 
    Total Degree Centrality & 5.54 & 2.96E-8 & 3.55E-7 & Yes \\ \hline 
    \end{tabular}
    \caption{Comparison of Network Metrics of Bots vs Humans}
    \label{tab:network_metrics}
\end{table}

\vspace{-1.5cm}

\section{Star Network Motifs}
\vspace{-0.2cm}

The singular ego node $n_0$ is one drawn from the set of \{bot,human\}. $n_0$ has multiple alters $n_a=\{n_{a1},n_{a2}...\}$, mainly three scenarios of alters: \{bot, human, mixed\}. Therefore, there are $2\times3=6$ primary combinations of ego-alters.

\begin{figure*}[h!]
    \centering
    \includegraphics[width=\linewidth]{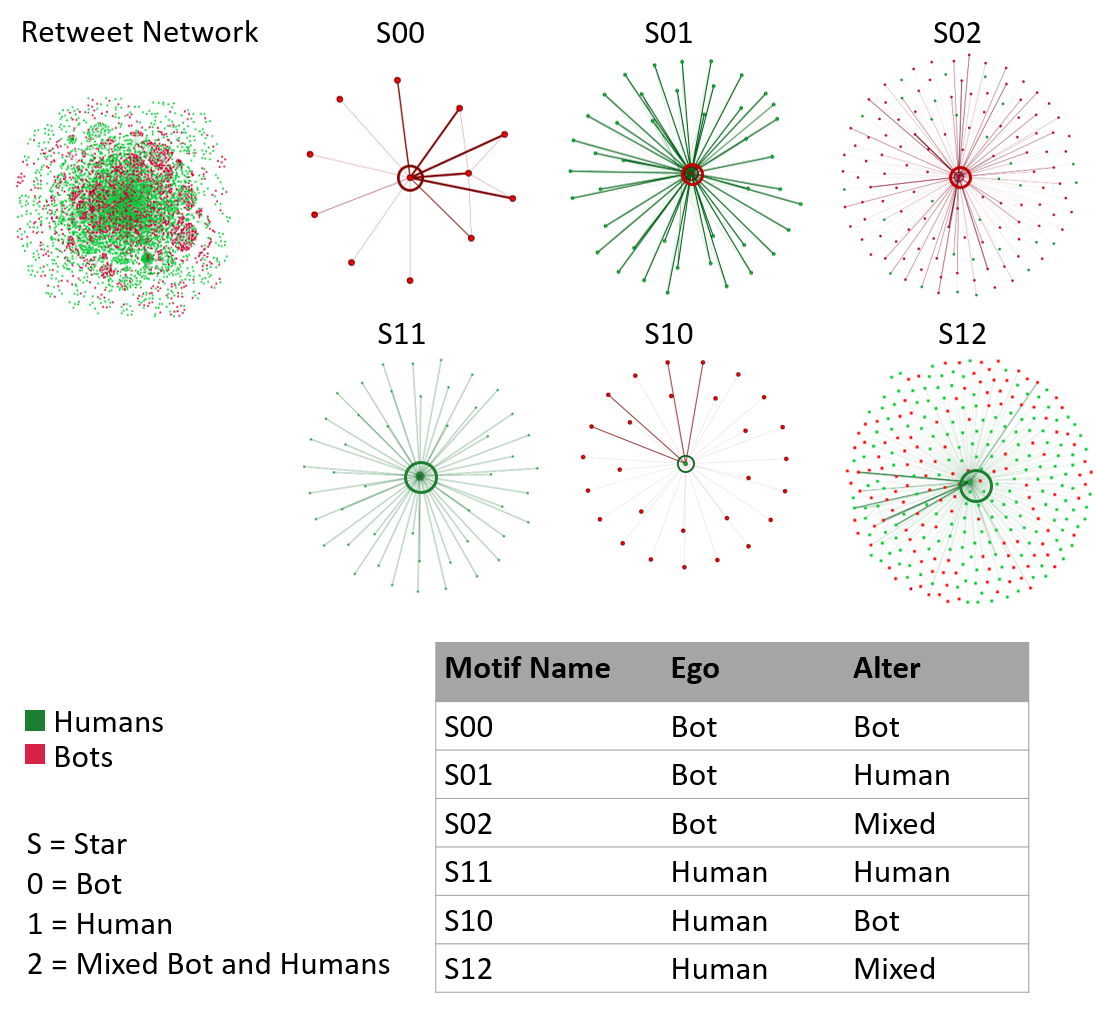}
    \caption{Variations of undirected star shaped networks}
    \label{fig:summary}
\end{figure*}

We thus create a notation to represent the variations of the undirected star motifs. The notation contains three characters $Sab$. $S$ indicates the star shaped network. $a={0,1}$ indicates whether the ego $n_0$ is a human or bot. $b={0,1,2}$ indicates the ordinal formulation of whether the alters $n_a$ are entirely bots, entirely humans, or a mixture of bots and humans.
The six star motifs are represented in \autoref{fig:summary}. Below, we describe each motif in its network structural form, and also the manifestation of each star motif in our dataset. Some of these accounts are still active, so usernames are redacted to preserve user privacy.

\textit{Motif S00.} $n_0$ is a bot, and $n_a$ are bots. The ego is an anti-vaccine bot that rejects and criticizes the COVID-19 vaccination program in the UK and the government's handling of the pandemic. The alters are other bots geolocated in the UK that supported the anti-vaccine stance.

\textit{Motif S01.} $n_0$ is a bot, and $n_a$ are humans. The ego has a conservative political ideology and frequently posts Bible quotes to support his ideology. Besides retweeting the conservative tweets, the posts that the alters authored also mock identity politics, particularly the pronouns and transgender communities. The alters also share posts with strong nationalist sentiments.

\textit{Motif S02.} $n_0$ is a bot, and $n_a$ are a mixture of humans and bots. The ego has a persona of a Japanese with a PhD, and has multiple social media presence, on X, Note, LinkedIn and Twilog. The ego has rapid posts spaced by seconds, which led us to agree with the bot algorithm's analysis. The alters are retweeting the ego as an authority figure.

\textit{Motif S11.} $n_0$ is a human, and $n_a$ are humans. The ego hosts a blogspot that publishes a collection of commentaries and opinion pieces. The topics focus on protests and unrest and feature images reminiscent of political cartoons. The posts by the ego turned into a community-maintained spread, where the alters are mostly South Americans and Argentinians. Four alters belong to prominent Argentine lawyers and politicians, and frequently retweet the ego (thicker lines). Post-analysis, the ego's account was terminated.

\textit{Motif S10.} $n_0$ is a human, and $n_a$ are bots. The ego is a news channel from India, and the alters were bot accounts associated with Mumbai neighborhoods. The ego regularly posts right-wing Hindu nationalist or anti-Muslim content, and alters associated with entire neighborhoods provide large potential audience for their messaging and rhetoric.

\textit{Motif S12.} $n_0$ is a human, and $n_a$ are a mixture of humans and bots. The ego is the BBCNews X account, a news network that reports on international news. The account is operated by a team of human social media managers. The alters are a mixture of bot and human agents that retweet the news.



\section{Discussions and Conclusions}
The star network motifs identified in this study are representative of basic building blocks of information amplification within the social media ecosystem that comprises of bot and human agents. These are emergent local structures formed from the intentional retweeting efforts of the agents. Our analysis of motifs within the retweet communication network reveals how they serve as mechanisms for content dissemination and influence propagation during a crisis event such as the COVID-19 pandemic. This includes spreading political ideologies, purporting anti-vaccination campaigns, echoing authorities, and disseminating news.

The six primary star motif patterns demonstrated distinct amplification strategies employed by different agent combinations. Bot-ego motifs (S00, S01, S02) exhibited characteristics of deliberate, and perhaps coordinated, information spreading, where bots leverage their central network position to rapidly disseminate content to multiple peripheral nodes. This aligns with previous research that show how bots serve as information super-spreaders and create efficient information propagation pathways \cite{broniatowski2018weaponized}. Conversely, human-ego motifs (S10, S11, S12) demonstrate more organic community formation patterns with natural clustering around authoritative sources and shared interests, an observation that is consistent with the principle of homophily \cite{mcpherson2001birds}. The motifs with mixed agent alters (S02, S12) reveal the complex interaction dynamics in a social network, and therefore bot detection and influence mitigation strategies must account for these hybrid communication patterns.

Star motifs contribute to social cybersecurity research \cite{carley2020social} by providing a structural framework for understanding how influence operations can manifest at the network level. The S00 motif represent a pattern of coordinated inauthentic behavior through a co-retweet network, and could be identified in monitoring systems to trigger alerts when multiple of such patterns emerge around sensitive topics like politics \cite{ng2023combined}. The S01 motif represent a pattern where automated agents can influence genuine human users, consistent with research on social influence and peer effects of bots \cite{ng2022pro}. The S10 motif demonstrate how authentic human content creators can be artificially amplified through sets of bots, potentially distorting the perceived popularity of their messages \cite{santini2020making}.

\paragraph{Strengths and weaknesses of star network motifs} 
From an individual perspective, the ego has high influence and visibility, greater access to information, and may receive more favors. The hub-and-spoke architecture enables efficient information dissemination, allowing the ego to simultaneously reach multiple alters with minimal network traversal. This makes them highly effective for broadcast communication, evident in S12, where the ego is a news channel that rapidly distributes international news updates to both bots and human agents. Similarly, in S01, the bot ego effectively influences human alters to share nationalist content. Star motifs create clear authority structures where alters receive information from a single trusted ego, reducing information ambiguity that arises from multiple conflicting sources, as demonstrated in S11, where the human ego has a coherent community of followers that retweet its authoritative pieces.

However, this centralized position comes at a potential cost of high emotional and cognitive load, being overly depended upon, and being a potential bottleneck for information flow. This makes the ego a single point of failure, because the entire communication structure is dependent upon it. This network fragility is seen in the S11 variation, where the ego's account termination disrupts the flow of political commentaries to the network formed around the node. The star motif also creates content control points that limit the diversity of information flow. Alters are primarily dependent on the ego for information, potentially creating echo chambers. In the S00 motif, the anti-vaccine bot ego spurs a network of bot alters that reinforces shared narratives. 

From a group perspective, stars are efficient communication structures with clear leadership and conflict mediation, but peripheral actors may feel isolated . The precise manifestation of these group dynamics varies based on agent composition. In homogeneous configurations like S00, the star structure facilitates synchronized messaging and collective amplification of share anti-vaccine narratives. Conversely, mixed-agent configurations like S12 illustrate how the star structure can serve diverse audiences simultaneously, where both human and bot alters consume and redistribute information from the ego to their respective networks. However, when used in a group, the star highlights vulnerabilities in information integrity. The S10 motif shows where human news channels are artificially amplified through bot networks, potentially distorting perceived public engagement and creating false impressions of grassroots support.

\paragraph{Limitations and Future Work} Our study is not without limitations. Firstly, our data was collected using selected relevant hashtags with the X Streaming API, and therefore would only encompass the tweets returned by the API. Secondly, our study mainly looked at undirected star shaped motifs. This calls for further in-depth studies to directed motifs, because directionality layers the flow of information diffusion and the trend of information amplification. Studying directed network graphs can facilitate downstream analysis to uncover patterns of strategic communication targets and infer possible intents. 

One current direction of expansion is to develop quantitative tools to identify the star network motifs at scale, and within large volume of data.
Finally, star-shaped motifs are not the only types of motifs that occur within social media conversational data. Future work include the studies of other motifs formations and their meanings within social media communication networks.

\paragraph{Conclusions}
This study provides insights towards the network structures of information amplification through the lens of star network motifs. By analyzing the bot-human retweet pattern within this motifs, we illustrate how different agent combinations created distinct pathways for influence propagation. 
Our taxonomy of six primary star motif patterns offers a systematic framework for understanding how organic and inorganic agents interact within X during the COVID-19 pandemic. This offers a structural perspective that complements past work focused on individual agent characteristics as ingredients for influence spread \cite{broniatowski2018weaponized,chang2022comparative}.
This work provides a foundation for future research into the structural dynamics of online influence, and builds an initial framework to understand one of the fundamental network patterns that enables information amplification.


%
%
%
\bibliographystyle{splncs04}
\bibliography{references}

\end{document}